\documentclass[10pt,twocolumn]{article}
\usepackage{ol2}

\usepackage{amsmath}
\usepackage{graphicx}
\usepackage{subfigure}
\usepackage{verbatim}

\begin{document}
\twocolumn[
\title{Picoradian deflection measurement with an interferometric quasi-autocollimator using weak value amplification}

\author{Matthew D. Turner,$^*$ Charles A. Hagedorn, Stephan Schlamminger,$^1$ Jens H. Gundlach}

\address{
Center for Experimental Nuclear Physics and Astrophysics, University of Washington, Seattle, Washington, 98195, USA\\
$^1$Currently with the National Institute of Standards and Technology, Gaithersburg, Maryland 20899, USA\\
$^*$Corresponding author: mdturner@phys.washington.edu}

\begin{abstract}We present an ``interferometric quasi-autocollimator'' that employs weak value amplification to measure angular deflections of a target mirror.  The device has been designed to be insensitive to all translations of the target.   
We present a conceptual explanation of the amplification effect used by the device.  
An implementation of the device demonstrates sensitivities better than 10 picoradians per root hertz between 10 and 200 hertz.  
\end{abstract}


]

\noindent Weak value amplification was first posited by Aharanov \emph{et al.} in 1988 \cite{aharanov}.  Using this phenomenon, the value of a measurement can be effectively amplified by proper preselection and postselection of the ensemble of particles used to make the measurement.  
Experimental realization of weak value amplification was demonstrated by Ritchie \emph{et al.} in 1991 \cite{ritchie}, and has since been used in a number of demonstrations and experiments, including the first measurement of the photonic spin Hall effect \cite{hosten}.  Aharanov \emph{et al.} have suggested a time-symmetric quantum mechanics formalism to simplify the explanation of this and other pre- and postselection effects \cite{physicstoday}.  

\begin{figure}
 \centerline{\subfigure[]{\label{rochester}\includegraphics[width=8.4cm]{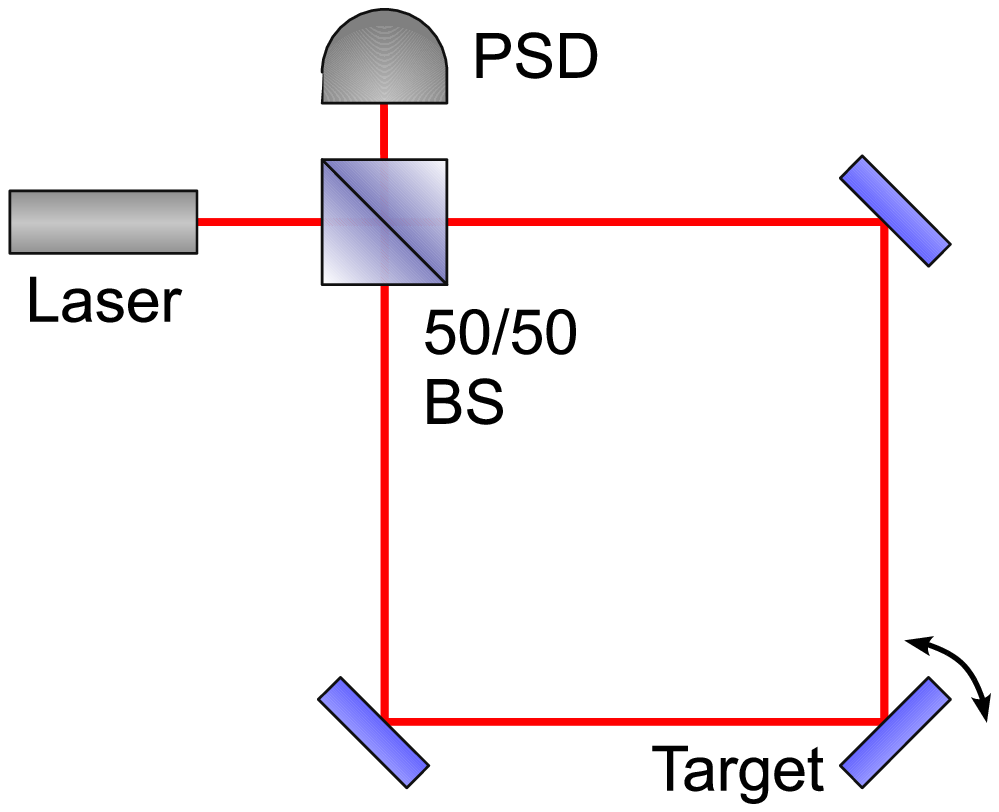}}}
 \centerline{\subfigure[]{\label{schematic}\includegraphics[width=8.4cm]{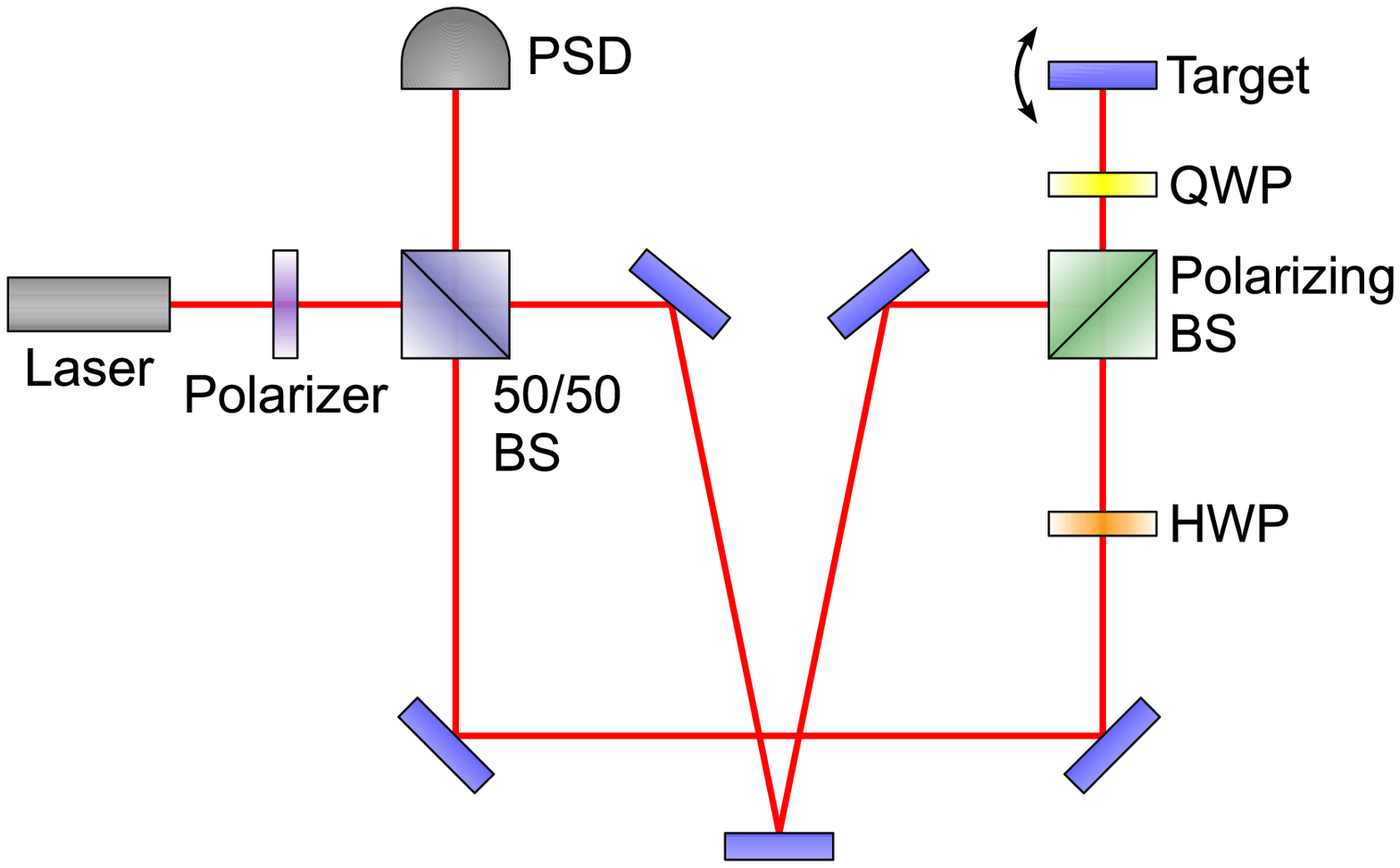}}}
  \caption{\label{schematics}(Color online) (a) Schematic diagram of the Rochester design.  A phase offset between the two optical paths is created either by the insertion of a Soleil-Babinet compensator or an out-of-plane deflection of one of the mirrors.  (b) Schematic diagram of the interferometric quasi-autocollimator (iQuAC).  The beam polarization is used to separate the two paths after reflection from the target.  In both setups, an in-plane angular deflection of the target mirror results in a proportional displacement of the beam spot on the position-sensitive photodetector (PSD).  }
\end{figure}

A recent paper by Dixon \emph{et al.}, of the University of Rochester, describes a device that uses weak value amplification to measure angular deflection \cite{dixon}.  
Their device, shown schematically in Fig. \ref{rochester} and referred to here as the Rochester design, utilizes a Sagnac-interferometer geometry.  
The introduction of a path-dependent phase offset, via either a half-wave plate and Soliel-Babinet compensator or a small out-of-plane deflection of one of the mirrors (see \cite{starling}), allows photons to exit the nominally dark port.  In-plane deflections of the target give the exiting photons a path-dependent transverse momentum.  For deflections resulting in a transverse momentum smaller than the transverse momentum uncertainty, weak value amplification results in proportional displacements of the beam spot at the dark port.  The amount of amplification can be adjusted by varying the phase offset between the two paths.  

The ability to measure small angular deflections quickly, precisely, and accurately is important in many fields of experimental physics.  
In the E\"ot-Wash experimental gravity group at the University of Washington, our torsion balance experiments depend upon our ability to measure minuscule angular deflections \cite{todd, prd}.  The most commonly used measurement device is an optical autocollimator, which collimates the light of a point source, reflects it off a target mirror, and then focuses the light onto a position-sensitive photodetector using the original collimating lens.  Autocollimators are insensitive to displacements of the target, eliminate the effect of optical aberrations in the lens, and average over the reflecting surface of the target.  Our best autocollimators have a sensitivity of $\sim1\ \mathrm{nrad}/\sqrt{\mathrm{Hz}}$.  
Similar angular deflection measurements can also be made using homodyne interferometers that compare the path length of beams incident on two separate locations on the target.  

The Rochester design offers the same intrinsic quantum noise (i.e. photon shot noise) limit as an autocollimator, as explained by Starling \emph{et al.} \cite{starling}.    An autocollimator achieves its reduced quantum noise, as compared to a simple beam-deflection setup, because the size of the beam spot on the detector is reduced by focusing the beam.  The enhanced signal-to-quantum-noise ratio of the Rochester design, however, results from amplification of the signal.  Thus the ratio of the signal to technical noise (e.g. electronic or digitization noise) is also enhanced.  
The Rochester design has two other specific advantages over an autocollimator.  The design is size-independent---the beam spot displacement and device sensitivity do not depend on the size of the setup.  Additionally, the number of photons incident on the detector for a given signal-to-quantum-noise ratio is reduced, allowing for the use of low-saturation-intensity detectors.  A homodyne interferometer has the same intrinsic quantum noise limit as an autocollimator and the Rochester design \cite{putman}, but the Rochester design is less sensitive to optical-path-length fluctuation and  uses a position-based, rather than intensity-based, measurement.  

Besides being free to rotate about the axis of the torsion fiber, torsion pendulums also swing.  If the Rochester setup were used to monitor a torsion balance, displacements normal to the mirror surface would result in equal displacements (multiplied by $\sqrt{2}$) of the laser spot on the detector.  Such displacements would be indistinguishable from rotation of the torsion pendulum.  Despite the implementation of swing-damping techniques, our torsion balances have displacement noise amplitudes of about $5\ \mathrm{\mu m}$, which would limit angular detection using the Rochester setup to sensitivities well above those available with our existing autocollimators.  

To make the benefits of the Rochester device available for use as an autocollimator replacement, we have developed an angle-measuring device designed to be insensitive to translations and out-of-plane angular displacements of the target mirror.  We refer to this device as an interferometric quasi-autocollimator (iQuAC).  In order for the device to be insensitive to target displacements, the laser beams incident on the target must be approximately normal to the reflecting surface.  This requires an extra degree of freedom by which the two paths of the Rochester design can be distinguished, which is provided by manipulating the polarization of the beams.  

In one form of our design, shown in Fig. \ref{schematic}, the light passes through a polarizer and into a 50/50 beam splitter.  The reflected light then reflects off two mirrors and passes through a half-wave plate, rotating its polarization by 90 degrees and allowing it to pass through a polarizing beam splitter.  It then passes through a quarter-wave plate, reflects off the target mirror, and returns through the quarter-wave plate.  The two passes through the quarter-wave plate rotate the polarization by another 90 degrees, causing the light to reflect off the polarizing beam splitter.  The light then is reflected by three mirrors and is returned to the 50/50 beam splitter.  Light that initially passes through the 50/50 beam splitter follows the same path in reverse.  An intentional slight out-of-plane misalignment of one of the non-target mirrors allows some light to exit the nominally dark port of the interferometer due to the resulting difference in path lengths.  An in-plane rotation of the target results in a displacement of the beam spot at the dark port, and the magnitude of the out-of-plane misalignment controls the weak value amplification factor of the displacement.  The mirror arrangement in the interferometer is such that the two paths between the two beam splitters are of equal length, fixing the target in the center of both paths between the light and dark ports and mitigating the effects of out-of-plane angular displacements of the target.  To allow the weak value amplification to occur, the number of reflections in each path is such that one path has an even number of reflections after the target and the other has an odd number, resulting in the light from the two paths having opposite in-plane transverse momentum when exiting the dark port.  

\begin{figure}
\centerline{\includegraphics[width=\linewidth]{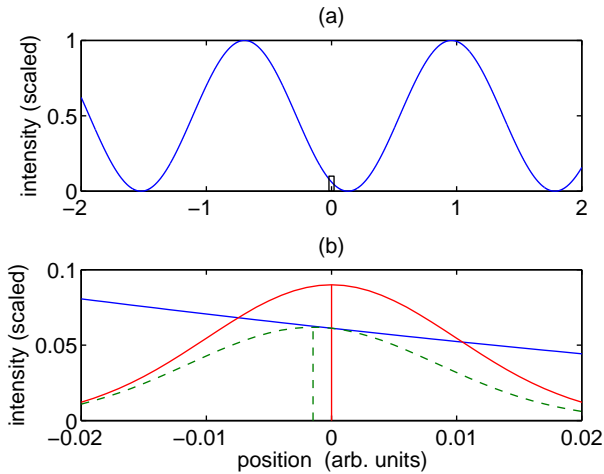}}
\caption{\label{explanation} (Color online) (a) An example of the fringe pattern which would result from a spatially uniform source being directed into the iQuAC.  The central minimum is shifted due to a phase offset between the two paths.  (b) The portion of the previous plot denoted by the black box, with an example Gaussian profile (red, solid).  The result of multiplying the Gaussian with the fringe pattern appears as a shifted Gaussian profile (green, dashed).  The first Gaussian profile has been scaled for clarity.  } 
\end{figure}

Howell \emph{et al.} have provided mathematical derivations, both quantum-mechanical and classical, of the weak value amplification scheme used by the Rochester design and the iQuAC \cite{howell}.  The classical description can also be explained in conceptual terms.  Suppose a spatially-uniform coherent source is directed into the interferometer.  If the system were to be exactly aligned with zero phase offset between the two paths, complete destructive interference would occur at the dark port.  If the target were to be rotated in plane, a series of equally spaced fringes, oriented perpendicular to the plane of the apparatus, would appear.  The center point would have zero intensity, and the fringe spacing would be inversely proportional to the angle of the target.  The introduction of a phase offset between the two paths would result in the fringe pattern being displaced by the same phase, regardless of the fringe spacing (or angle of the target), so that the intensity of the center point is the same for any angular displacement of the target.  A representation of these fringes is shown in Fig. \ref{explanation}(a).  

Now suppose that a Gaussian-profile source is used.  The fringe pattern would still be present but multiplied by the beam profile.  
For sufficiently small angular deflections, the fringe spacing would be large compared to the width of the beam, and the fringe pattern in the region of the beam spot could be approximated by a linear function.  
A linear function of a small slope multiplied by a zero-centered Gaussian is approximately equal to a Gaussian of equal width that has been translated by some distance, i.e. for $b \sigma \ll a$,   
\begin{multline}\label{approxgaussian}
\left(a + b x\right) \exp\left[-x^2/\left(2\sigma^2\right)\right]\\
\approx a 
\exp\left[\frac{-\left(x-b\sigma^2/\! a\right)^2}{2 \sigma^2}\right].
\end{multline}
For sufficiently wide fringe spacing (or small angular deflections), the fringe pattern in the region of the Gaussian can be approximated by the linear function 
$
\sin^2(\phi/2) - 2 k_0 \theta x \sin \phi,
$
 where $\phi$ is the phase offset, $k_0$ the wave number, $\theta$ the angular deflection of the target, and $x$ the transverse position.  Replacing the values in \eqref{approxgaussian} and approximating for small values of $\phi$ gives a translation of $8 k_0 \sigma^2 \theta/\phi$.  This result corresponds to an amplification, as compared to a simple beam-deflection setup with target-detector separation $l_{td}$, of $4 k_0 \sigma^2/(\phi\, l_{td})$, and matches the result of Howell \emph{et al}.  A representation of the multiplication of the fringe pattern with a Gaussian profile is shown in Fig. \ref{explanation}(b).  

\begin{figure}
\centerline{\includegraphics[width=\linewidth]{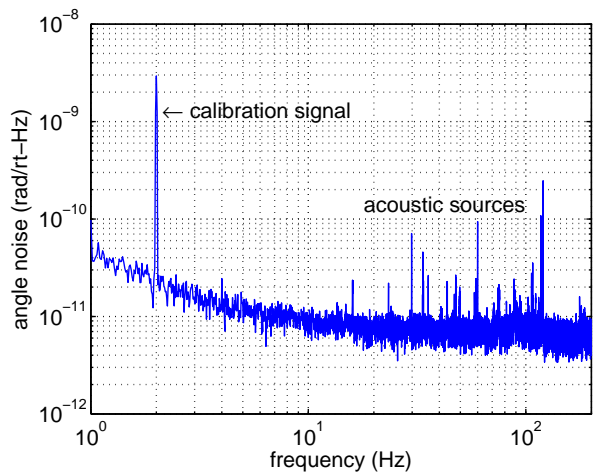}}
\caption{\label{noise}(Color online) A plot of the noise floor of an implemented iQuAC setup.  A 620-prad calibration signal is visible at 2 Hz.  A weak value amplification factor of 60, as compared to a simple beam-deflection measurement, was measured for this data set.  }
\end{figure}

We have constructed an iQuAC setup using standard optics, a 5-mm position-sensitive photodetector, and a 10-mW, 660-nm diode laser connected via a single-mode fiber to a 3.4-mm diameter collimator.  The signal from the photodetector is amplified, low-passed, and then digitized and read using a data-acquisition board.  A piezoelectric disk connected to a signal generator is used to generate in-plane angular displacements of one of the mirrors.  It is calibrated by removing the 50/50 beam splitter and making a simple beam-deflection measurement.  A 2-Hz sine wave voltage signal applied to the disk is used as a calibration signal to determine the weak value amplification factor, which we have observed to be in the range of 40 to 100.  A noise plot from the setup is shown in Fig. \ref{noise}.  We observe a noise floor of less than $10\ \mathrm{prad}/\sqrt{\mathrm{Hz}}$ in the 10-200 Hz band, independent of amplification.  Below 10 Hz, $1/f$ noise dominates, and above 200 Hz, acoustic pickup increases.  We plan to construct a monolithic version of the device to reduce these effects.  

This work was supported by NSF grants PHY0653863 and PHY0969199, NASA grant NNX08AY66G, 
and by DOE funding for the Center for Experimental Nuclear Physics and Astrophysics.


\begin{thebibliography}{10}
\newcommand{\enquote}[1]{``#1''}

\bibitem{aharanov}
Y.~{Aharonov}, D.~Z. {Albert}, and L.~{Vaidman}, Physical Review Letters
  \textbf{60}, 1351 (1988).

\bibitem{ritchie}
N.~W.~M. Ritchie, J.~G. Story, and R.~G. Hulet, Phys. Rev. Lett. \textbf{66},
  1107 (1991).

\bibitem{hosten}
O.~{Hosten} and P.~{Kwiat}, Science \textbf{319}, 787 (2008).

\bibitem{physicstoday}
Y.~Aharonov, S.~Popescu, and J.~Tollaksen, Physics Today \textbf{63}, 27
  (2010).

\bibitem{dixon}
P.~B. Dixon, D.~J. Starling, A.~N. Jordan, and J.~C. Howell, Phys. Rev. Lett.
  \textbf{102}, 173601 (2009).

\bibitem{starling}
D.~J. Starling, P.~B. Dixon, A.~N. Jordan, and J.~C. Howell, Phys. Rev. A
  \textbf{80}, 041803 (2009).

\bibitem{todd}
S.~Schlamminger, K.-Y. Choi, T.~A. Wagner, J.~H. Gundlach, and E.~G.
  Adelberger, Phys. Rev. Lett. \textbf{100}, 041101 (2008).

\bibitem{prd}
G.~L. Smith, C.~D. Hoyle, J.~H. Gundlach, E.~G. Adelberger, B.~R. Heckel, and
  H.~E. Swanson, Phys. Rev. D \textbf{61}, 022001 (1999).

\bibitem{putman}
C.~A.~J. {Putman}, B.~G. {de Grooth}, N.~F. {van Hulst}, and J.~{Greve},
  Journal of Applied Physics \textbf{72}, 6 (1992).

\bibitem{howell}
J.~C. Howell, D.~J. Starling, P.~B. Dixon, P.~K. Vudyasetu, and A.~N. Jordan,
  Phys. Rev. A \textbf{81}, 033813 (2010).

\end{thebibliography}
\end{document}